\documentclass[twocolumn]{revtex4}
\usepackage{graphicx}
\begin{document}
\title{A toy model for quantum mechanics}
\author{S.J. van Enk}
\affiliation{Department of Physics, University of Oregon\\
Oregon Center for Optics and
Institute for Theoretical Science\\
 Eugene, OR 97403\\}
\begin{abstract}
The toy model used by Spekkens [R.~Spekkens, Phys. Rev. A {\bf 75}, 032110 (2007)] to argue in favor of an epistemic view of quantum mechanics is extended
by generalizing his definition of pure states (i.e. states of maximal knowledge) and by associating measurements with all pure states.  The new toy model does not allow signaling but, in contrast to the Spekkens model, does violate Bell-CHSH inequalities. Negative probabilities are found to arise naturally within the model, and can be used to explain the Bell-CHSH inequality violations. 
\end{abstract}
\maketitle
\section{Introduction}
In recent years various foundational principles have been put forward for quantum mechanics that are concerned with {\em information}  \cite{hardy1,hardy2,hardy3,bub,fuchs}.
The hope is to formulate simple principles from which one can derive quantum mechanics, much as special relativity can be derived from the principle of relativity: that the laws of physics are the same in any inertial reference frame.

A concrete example of such a principle \cite{zeilinger} relevant to the present paper is that an elementary system carries one bit of information, i.e., the truth value to a single binary proposition. This principle allows one, according to Zeilinger \cite{zeilinger}, to explain the randomness in quantum mechanics when considering measurements corresponding to {\em other}, different propositions. However, Timpson \cite{timpson} pointed out that Zeilinger's principle does not lead to quantum mechanics. That objection seems valid: after all, a classical bit certainly satisfies the above definition of an elementary system.
Moreover, one must be careful here to distinguish information encoded in a system, information needed to describe a system, and information obtained from measuring a system. Whereas in classical physics the three may well be (in principle if not in practice) the same, in quantum mechanics they are certainly not \cite{cavesfuchs}.

Spekkens \cite{spekkens} added a crucial ingredient to Zeilinger's principle. He showed that many features of quantum mechanics can be reproduced when requiring that an elementary system, in addition to answering one binary question, also leaves one binary question  {\em un}answered. More precisely, Spekkens presented a toy model in which physical states describe partial knowledge about hidden variables. That is, physical states correspond to {\em epistemic} states (states of knowledge) in which the knowledge
one has about the values of quantities describing the underlying {\em ontic} state (a state of reality) is limited.
His ``knowledge balance principle'' states that the number of questions that are answered in a physical state can be at most equal to the number of questions left unanswered.  In a state of maximal knowledge (which will be called {\em pure states} here \footnote{In our toy model this conforms to standard terminology: the set of allowed epistemic states is a convex set, and maximal-knowledge states are the extreme points of the set, hence pure. In the Spekkens toy model, on the other hand, the set of epistemic states is actually not convex. Perhaps this observation gives an additional reason for extending the set of allowed epistemic states beyond the Spekkens model.}), one knows the answers to exactly {\em half}
of the total number of questions necessary to specify the ontic state. 

If we assume the ontic state is fully described by an even number of variables each taking on one of a fixed number $p$ values, then for a pure state one knows the values of exactly half of those variables. 
For $p=2$ one knows one bit of information and thus the Spekkens model
includes and extends the above-mentioned foundational principle proposed by Zeilinger. The main point of the (purely classical) Spekkens model is that it reproduces many features of quantum mechanics usually considered nonclassical. For instance, the toy model describes teleportation, superdense coding, a no-cloning theorem, and the noncommutativity of measurements, just to name a few \cite{spekkens}. But one quantum feature {\em not} reproduced by the model is violations of Bell-CHSH inequalities \cite{bell,chsh}.

We will consider here a natural generalization of that toy model by allowing pure states to correspond to partial knowledge about more than half of the variables. More importantly, we will associate a measurement and an observable with every pure state.
A pure state will still correspond to having ``half the information'' needed to specify the ontic state, and various measures of information will be considered to specify what exactly ``half the information'' means. The main point is that this extension turns out to reproduce even more salient features of quantum mechanics, in particular violations of the Bell-CHSH inequalities.
The extension actually connects the Spekkens toy model
with "generalized probability theories" as considered in Refs. \cite{hardy1,hardy2,hardy3,barrett,barnum}. The latter theories consider probabilities for measurement outcomes, while in Spekkens model the probabilities refer to knowledge about an underlying state of reality. The correspondence between the two probability distributions, or the lack thereof, is in fact crucial for the present toy model.
\section{Toy model}
\subsection{Observables and measurements}
In our toy model an {\em elementary} system is described by two independent variables $X_a$ and $X_b$, that each can take on $p$ different values, where $p$ is a prime number \footnote{We can certainly allow composite numbers but then the system described can be considered to be a composite rather than elementary system.}. 
Those systems will be referred to as ``elementary systems of type $p$.'' 

The total number of different values taken by the two variables is then $N=p^2$ (and this is also the number of distinct ontic states).
The values taken on by the variables $X=(X_a,X_b)$ are denoted by
$x=(x_a,x_b)$,
with
\[
x_{a,b}\in \{0,1,\ldots p-1\}.
\]
More precisely, we will assume henceforth the values are taken
in $F_p$, the field of integers modulo $p$, equipped with multiplication and addition.
For the moment we define a {\em pure state} to be a state where we know the value of one particular variable and nothing else. This variable could be $X_a$ or $X_b$, or more generally
any variable of the form 
\[
k_aX_a+k_bX_b,
\]
with $k_{a,b}\in F_p$. Thus we restrict ourselves  to ``observables'' that are linear functions of $X_a$ and $X_b$.
We have $p^2-1$ nontrivial observables
that can take on $p$ different values, and the trivial observable $\equiv 0$.

Now the observables that we defined are in fact not all different, as we can always divide out a nonzero factor $k_a$ or $k_b$, to write either
\[
k_aX_a+k_bX_b=k_a (X_a+k_b k_a^{-1}X_b)\, {\rm for }\, k_a\neq 0,
\]
or 
\[
k_aX_a+k_bX_b=k_b (k_a k_b^{-1}X_a+X_b)\, {\rm for }\, k_b\neq 0.
\]
Thus we can use as different nontrivial observables the set
\begin{equation}\label{O}
{\cal O}=\{X_0=X_a; X_{k+1}=X_b+kX_a,\,{\rm for}\,k=0\ldots p-1\},
\end{equation}
whose size is $R=|{\cal O}|=p+1$. 
In a pure epistemic state exactly one value of the observables in ${\cal O}$ is specified. Thus there are $p(p+1)$ different pure epistemic states.

These $p+1$ observables $X_i$ for $i=0\ldots p$ are mutually unbiased, in that
knowledge of one such variable tells one {\em nothing}
about the value of another variable. This implies, for instance, that we can assign any pair of observables out of the set ${\cal O}$ to represent the ontic state.
In other words, the space of observables (not of states!) is a 2-D vector space over the field $F_p$, and in it we can distinguish $R=p+1$ different rays (vectors up to an overall irrelevant factor) that act as different (but obviously not linearly independent) observables.

For each observable there are measurements associated with it:
The allowed measurements in the toy model answer a question of the form ``is the value of the variable $X_i$ equal to $x_i$?''.
This model is the algebraic formulation of the Spekkens model for $p=2$ and generalizes it for other prime numbers $p$. In the Appendix we will present the graphical representation of Spekkens' toy model, including the present extension of the model.
\subsection{Measures of information and pure states}
So far we have defined pure states as states for which we know the value of exactly one observable. These pure states are identical to the pure states defined in \cite{spekkens} for $p=2$. Let us call these pure states the {\em canonical} pure states of a system of type $p$. 

We can extend the definition of pure states, as follows.
Suppose we assign to a system of type $p$ a probability distribution for the possible values of 2 observables, say, $X_a$ and $X_b$, denoted by
$P(x_a,x_b)$. Now the amount of information we have can be quantified for any such function $P$ provided we use some particular measure of information (or perhaps we should say ``predictability'' rather than ``information'').
Such measures can be constructed from reasonable axioms, discussed in, e.g., \cite{uffink}. The axioms single out a class of measures $M_r$ of predictability, corresponding to Schur-convex functions, parametrized by 
a real number $r$,
\begin{equation}\label{M}
M_r(P)=\left(\sum_x P(x) (P(x))^r\right)^{1/r},
\end{equation}
where the sum is over all possible values of $x=(x_a,x_b)$.
Here we need $r>-1$ in order for $M_r$ to be strictly Schur convex \cite{uffink}. Strict Schur convexity ensures that predictability increases
when the probability distribution is ``more concentrated.''

If we could know the value of {\em both} variables $X_a$ and $X_b$, we would have $P({x_0})=1$ for one particular value $x=x_0$ with all other probabilities being zero. We would thus have $M_r(P)=1$. Knowing only one variable, but nothing about the other variables, corresponds to a probability distribution where $P(x)=1/p$ for $p$ different values $x$ and zero for the remaining ones.
Then we have $M_r(P)=1/p$. This value corresponds to a pure state in our toy model. To see in what sense this value is  ``halfway'' between full knowledge and no knowledge at all, consider that $M_r(P)=1/p^2$ when we have $P(x)=1/p^2$ for all $x$. [So we could use $\log_2(M_r(P))$ as an additive measure of predictability.]  
Note that the value of $M_r(P)$, for any value of $r$, does not depend on having singled out the observables $X_a$ and $X_b$ rather than some other pair from the set ${\cal O}$, as choosing a different pair merely corresponds to relabeling the values of the observables. 

We now define a {\em pure state} to correspond simply to {\em any} probability distribution $P$ such that $M_r(P)=1/p$. Thus we have the following crucial rule of our toy model:
\begin{equation}
{\rm Pure\, state\, of\, 1\, system\, of\, type\,}p: \,\,M_r(P)=1/p.
\end{equation}
We have to make a choice for $r$ in order to make this definition unique. Indeed, in general a probability distribution $P$ that is pure for a particular value of $r$ is not pure for different values of $r$. The canonical pure states, on the other hand, are pure for any value of $r$.

There are two obvious choices for $r$: we could choose $r=1$, which would correspond to the measure of information advocated by Brukner and Zeilinger in \cite{brukner} as being most useful in quantum mechanics,  or we could choose $r\rightarrow 0$ in which case $M_r(P)=2^{-H(P)}$,
with $H(P)=-\sum_i p_i\log_2 p_i$ the Shannon entropy. At this point, though, we do not make any choice yet.
\subsection{An equivalent characterization of pure states}
We have defined pure states in terms of the joint probability distribution $P$ for the variables $X_a$ and $X_b$. There are $p^2-1$ independent probabilities $P(x_a,x_b)$. Alternatively, we could use the probabilities for the $p+1$ observables $X_i, i=1\ldots p+1$ in the set ${\cal O}$ to take on the different values $x_j\in F_p$. Let us denote those probabilities by $Q_i(x_j)$.
Each probability distribution $Q_i$ is determined by $p-1$ independent values $Q_i(x_j)$ for $j=0,\ldots p-2$, and so there are $(p-1)(p+1)=p^2-1$ independent probabilities $Q_i(x_j)$. This is the same number as the number of independent probabilitites $P(x_a,x_b)$. Indeed, one probability distribution can be expressed in terms of the other. We have, in particular, the definition of $Q_i$ in terms of $P$:
\begin{equation}
Q_i(x)=\sum_{x_a,x_b|x_i(x_a,x_b)=x}P(x_a,x_b).
\end{equation}
Here we define, in analogy to (\ref{O}),
\begin{eqnarray*}
x_0(x_a,x_b)&=&x_a; \\
x_{k+1}(x_a,x_b)&=&x_b+kx_a,\,{\rm for}\,k=0\ldots p-1.
\end{eqnarray*}
Conversely, we find by summing this relation over all $i$, using that $\sum_{x_a,x_b}P(x_a,x_b)=1$, and rearranging terms, 
\begin{eqnarray}\label{QtoP}
pP(x_a,x_b)&=&-1+\sum_i Q_i(x_i(x_a,x_b))\nonumber\\
&=&\sum_i \left\{Q_i(x_i(x_a,x_b))-\frac{1}{p+1}\right\}
\end{eqnarray}
So instead of using $M_r(P)$ as our measure of information to determine what pure states are, we might as well use the probability distributions $\{ Q_i\}$ and write
\begin{eqnarray}\label{PtoQ}
M_r(P)
&=&\left (
\sum_{x_a,x_b} \left[
\frac{1}{p}\sum_i \left\{Q_i(x_i(x_a,x_b))-\frac{1}{p+1}\right\}
\right]^{r+1}
\right)^{1/r}\nonumber\\
&:=&N_r(\{Q_i\}).\end{eqnarray} 
We could view the description of a pure state in terms of the probability distributions $\{ Q_i\}$ an instrumentalist description, as it only refers to quantities that can be measured. 
In fact, except for the restriction that $N_r(\{Q_i\})\leq 1/p$, the instrumentalist description corresponds to the representation of states used in \cite{hardy1,hardy2,hardy3,barrett,barnum} in terms of probabilities for outcomes of a {\em fiducial} set of measurements, where in our case the fiducial set is ${\cal O}$. 

While the transition from epistemic to instrumental states may be quite natural, the measure of information $N_r(\{Q_i\})$ looks artificial.
On the other hand, in the special case $r=1$ we get after some algebraic manipulations the simpler and perhaps more natural-looking relations
\begin{eqnarray}\label{PtoQ1}
M_1(P)&=&\frac{1}{p}\sum_i M_1(Q_i)-\frac{1}{p}\\
&=&\frac{1}{p} \sum_i\sum_x \left(Q_i(x)-\frac{1}{p}\right)^2
+\frac{1}{p^2}.
\end{eqnarray}
Thus in the case of $r=1$ we see that $M_r(P)$ is determined by the sum
over $i$ of the same measure $M_r$ applied to the probability distributions $Q_i$, $M_r(Q_i)$.

In discussing the merits of $M_1$ as a measure of information Timpson in \cite{timpson} noted there is no particular reason to sum
$M_r(Q_i)$ over $i$ for any $r$. However, we see here that for $r=1$ there is in  fact a good reason, it quantifies the information in the underlying epistemic probability distribution $P$, through the relation (\ref{PtoQ1}).
\subsection{Multiple systems and signaling}
Consider now $N>1$ systems of type $p$, denoted by a superscript$(n)$ where $n=1\ldots N$. We use the joint probability distribution $P(X_a^{(1)},X_b^{(1)},\ldots X_a^{(N)},X_b^{(N)})$ to define pure (epistemic) states, in a straightforward extension of the previous discussion.
Namely, we will require $M_r(P)=(1/p)^N$ for a pure state of $N$ systems of type $p$. 
In the simplest case this corresponds to knowing exactly the values of $N$ variables out of the $2N$ independent variables describing the $N$ systems.
Moreover, we must require that for each marginal probability distribution $P'$ describing some subset of $N'<N$ systems, we have
$M_r(P')\leq (1/p)^{N'}$, in order not to violate the knowledge balance principle
for the subsystems.
However, this is not the only restriction placed on pure states.

For example, suppose we could have a state of two systems of type 2, characterized by $X_a^{(1)}=X_a^{(2)}$ and $X_b^{(1)}=0$. If we just consider the amount of information we have about such a state, then we see we know two bits of information about the joint system $(1,2)$, we know one bit about system 1 alone, and nothing about system 2 alone. Without further restrictions this would be an allowed pure state of 2 systems. However, such a state was excluded by Spekkens in his toy model, and here we will disallow it as well. The reason is that a measurement on system 2 of the variable $X_a^{(2)}$ would give too much information about system 1: both $X_a^{(1)}$ and $X_b^{(1)}$ would be known to us. Alternatively, we could 
impose the condition that the variable $X_b^{(1)}$ would be randomized because of the measurement of $X_a^{(2)}$ in order to keep the knowledge balance intact. But the cure would be worse than the disease, as it would allow {\em signaling}: the randomization of the value of $X_b$ of system 1 could be detected (with 50\% probability at least), and would signal the fact that on system 2 (possibly located far from system 1) a measurement of $X_a$ was performed.
\subsection{Extending the set of observables and measurements}\label{ExtendOM}
From now on we only consider the simplest case $p=2$. Suppose we have the (instrumental) state $S$ characterized by the probabilities
\begin{eqnarray}\label{S}
S: Q_0(1)&=&0.9\nonumber\\
Q_1(1)&=&0.9\nonumber\\
Q_2(0)&=&0.8\nonumber\\
r&\approx& -0.147.
\end{eqnarray}
The value of $r$ was chosen so as to have $N_r(\{Q_i\})=1/2$, so that
$S$ is a pure state. The underlying epistemic state is given by the probabilities 
\begin{eqnarray}\label{SP}
P(1,1)&=&0.8\nonumber\\
P(0,1)&=&P(1,0)=0.1\nonumber\\
P(0,0)&=&0,
\end{eqnarray}
such that $M_r(P)=1/2$ for $r\approx -0.147$.
For a system in the state $S$
we can guess the values of two observables (namely $X_a$ and $X_b$) quite well, and we are slightly worse informed about the third observable ($X_a+X_b$).

Now we wish to define a measurement
corresponding to the pure state $S$, just as we associated measurements with the canonical pure states. Since $p=2$ we need to define a binary question. The question we allow as a valid measurement is simply this: ``is the system in state $S$ or not?'' It is quite natural to assume that the probabilities of getting the answer ``yes'' to this question for the canonical pure states can be read off from the definition of the state $S$: 
if we have a system in a pure state where we know the variable $X_i$  to be
equal to $x_i$, then we simply  declare the probability of finding the system in state $S$ be
$Q_i(x_i)$. The reason for this choice is that, conversely, the probability of finding the result $X_i=x_i$ when measuring the variable $X_i$ on a system in the state $S$ is given by the same number, $Q_i(x_i)$, by definition of the state $S$.

How should we define the state $S^{\perp}$ corresponding to the answer ``no, the system is not in the state $S$''?
Here is one natural way of defining the state $S^\perp$:
Suppose we start out with a state about which we know nothing. That is, we have a maximally mixed state described by $Q_i(x_i)=0.5$ for all $i$. Then suppose we
perform the measurement $S$ vs $S^\perp$, but we forget the outcome. Then we should still ascribe the same mixed state, as we did not gain any information. Thus, an equal mixture of the states $S$ and $S^\perp$ should be equivalent to the maximally mixed state.
This uniquely defines $S^\perp$ to be
\begin{eqnarray}\label{Sp}
S^\perp: Q_0(1)&=&0.1\nonumber\\
Q_1(1)&=&0.1\nonumber\\
Q_2(0)&=&0.2\nonumber\\
r&\approx& -0.147.
\end{eqnarray}
That is, we just complement the probabilities
$Q_i(x_i)\rightarrow 1-Q_i(x_i)$.
This can be accomplished by the mapping
$P(x_a,x_b)\rightarrow 1/2-P(x_a,x_b)$, or equivalently by 
$x_i\rightarrow 1-x_i$.
All this is easily generalized to arbitrary pure states for arbitrary values of $r$. 

We also require that a system in state $S$ will always answer ``yes'' to the question, ``are you in S?''. This assumption is not as innocent as it may appear: it leads to an infinite ontological excess baggage, as was explained by Hardy \cite{hardy2}.
\section{Violating Bell-CHSH inequalities}
Perhaps surprisingly, the toy model as we have defined it allows for violations of Bell-CHSH inequalities \cite{bell,chsh}.
A crucial role is played by the extended set of observables and measurements of Section~\ref{ExtendOM}.

For example, suppose we start out with a state of two systems in which we know
\[
X_a^{(1)}=X_a^{(2)}\]
and 
\[X_b^{(1)}=X_b^{(2)}.
\]
This describes a pure state of 2 systems, in which we know exactly 2 independent variables out of a total of 4. We know nothing about each system individually, but the two systems are perfectly (and maximally) correlated. 
As shown in \cite{spekkens} such a state has many properties in common with a maximally entangled state of two qubits. For example, (i) teleportation is possible with such states, (ii) although the state of the two systems together is pure, the reduced states of the two subsystems are completely random (``mixed''), and (iii) superdense coding is possible.
On the other hand, in the Spekkens toy model Bell-CHSH inequalitites cannot be violated with such a state. 

In addition to the pure state $S$ defined in Section~\ref{ExtendOM}
we need to define one more pure state and its associated measurement.
We define a state $S'$ and its orthogonal complement $S'^\perp$ by the probabilities
\begin{eqnarray}\label{S'}
S':
Q_0(1)&=&0.9\nonumber\\
Q_1(0)&=&0.8\nonumber\\
Q_2(1)&=&0.9\nonumber\\
S'^\perp:
Q_0(1)&=&0.1\nonumber\\
Q_1(0)&=&0.2\nonumber\\
Q_2(1)&=&0.1\nonumber\\
r&\approx& -0.147.
\end{eqnarray}
This state is a ``rotated'' version of the state $S$,
as it can be obtained from $S$ by rotating $X_b\rightarrow X_a\rightarrow X_a+X_b\rightarrow X_b$.
This state $S'$ and its orthogonal complement $S'^\perp$ define another binary measurement.

Now contemplate performing measurements of the variables $X_a^{(1)}$ and 
$X_b^{(1)}$ on system 1, and
measurements $S^{(2)}$ and $S'^{(2)}$ on system 2.
The joint probabilities of the various possible outcomes are easily calculated straight from the definitions (\ref{S})--(\ref{S'}). 
To see that such joint probabilities indeed can be defined, consider the following. If we first measure the variable $X_a^{(1)}$ on system 1 then system 2 will "collapse" into a state with the same value
for $X_a^{(2)}$ as measured for system 1. A subsequent measurement of $S^{(2)}$ or $S'^{(2)}$ is then found to have the answer "yes" with the probabilities determined by (\ref{S})--(\ref{S'}). On the other hand, if we first measure $S^{(2)}$ or $S'^{(2)}$ then system 1 is collapsed to one of the states $S$ or $S^\perp$, or $S'$ or $S'^\perp$ depending on the outcome of the measurement on system 2 (this follows from the fact the initial state is perfectly correlated and repeated measurements on system 2 will persist in yielding the same answer as the first time). The probabilities for the various outcomes of measurements of $X_a^{(1)}$ or $X_b^{(1)}$ again follow from (\ref{S})--(\ref{S'}).
Hence the order of the measurements does not matter, and a joint probability distribution for the outcomes can be defined. 

Moreover, because of the property that an equal mixture of $S$ and $S^\perp$ or of $S'$ and $S'^\perp$ is equal to the maximally mixed state implies that signaling is impossible.

Back to the Bell inequalities. Suppose, for concreteness, we find $X_a^{(1)}=1$. We then infer that $X_a^{(2)}=1$ as well, from the definition of the bipartite state.
Then, if we measure the observable $S^{(2)}$, we get the result
``yes'' with probability $Q_0(1)=0.9$, whereas if we measure $S'^{(2)}$, we get the answer ``yes'' with probability $Q_0(1)=0.9$.
If on the other hand, we measure $X_b^{(1)}$ on system 1 and find it to be, say, $X_b^{(1)}=1$ then we know $X_b^{(2)}=1$. We thus get
the outcome ``yes'' after a measurement of $S^{(2)}$ on system 2 with probability $Q_1(1)=0.9$ 
and the outcome $S'$ on system 2 with probability $Q_1(1)=0.2$.
We can now easily construct the standard Bell-CHSH inequality $|B|\leq 2$ by defining
\begin{eqnarray}\label{B}
B:=\langle X_a^{(1)}(1) S^{(2)}+
X_b^{(1)}(1) S^{(2)}+\nonumber\\
X_a^{(1)}(1) S'^{(2)}
-X_b^{(1)}(1) S'^{(2)}
 \rangle,
\end{eqnarray}
where the answer ``yes'' counts as 1 and ``no'' as -1.
For the particular states $S$ and $S'$ we have defined here we
find
$B_{r=-0.147}=3\times(0.9-0.1)-(0.2-0.8)=3$. The violation allowed by quantum mechanics is
$B_{QM}=2\sqrt{2}\approx 2.8284$, and so the toy theory with $r=-0.147$
allows a stronger violation of the Bell-CHSH inequality than does quantum mechanics: it violates Tsirel'son's inequality \cite{tsirel}.

We can numerically maximize the Bell-CHSH parameter $|B|$ as a function of the parameter $r$, by maximizing over the possible pure states $S$ and $S'$ in the scenario analyzed above. The result is displayed in Figure~1.
\begin{figure}
\includegraphics[width=3.1in]{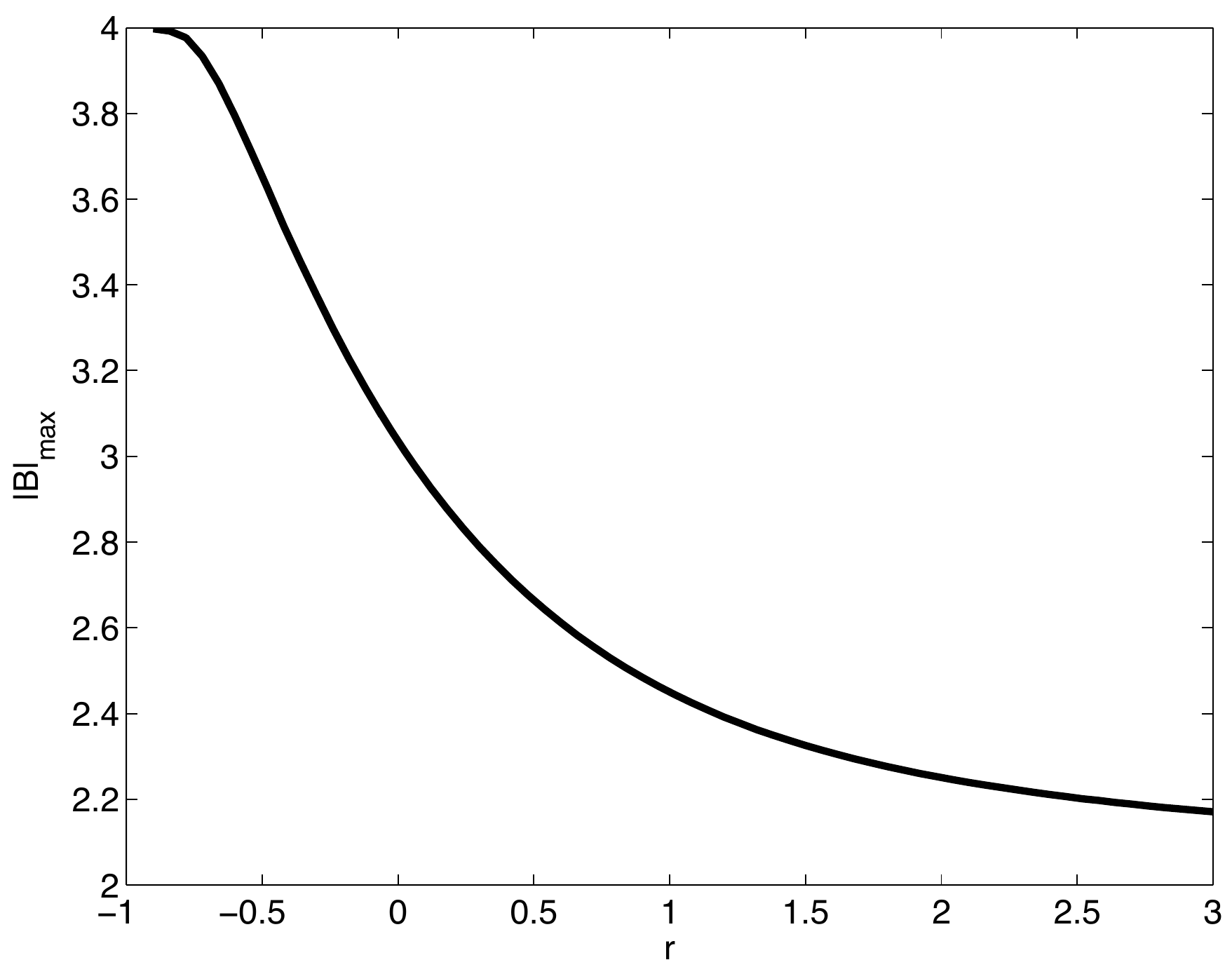}
\caption{Maximum violation of the Bell-CHSH inequality as a function of the parameter $r$ using measurements corresponding to pure states with $M_r(P)=1/2$.}
\end{figure}
For small values of $r$ the violation of the Bell inequality reaches its logical
limit $B_{{\rm max}}=4$, where as for large values of $r$ the violation becomes arbitrarily small, and $\lim_{r\rightarrow \infty} B_r=2$. All these toy models, then, violate Bell-CHSH inequalities but do not allow signaling. This is true even for the toy model corresponding to the limit $r\rightarrow -1$. This limit then mimics the so-called PR (Popescu-Rohrlich \cite{pr}) box, which allows the strongest possible correlations $|B|=4$ without allowing signaling.

But why can we violate a Bell-CHSH inequality in the toy model at all? Isn't it still a local hidden-variable model? The answer is no, because the states $S^{\perp}$ and $S'^{\perp}$ actually do not correspond to
valid epistemic states! Namely, the underlying probability distribution $P$ for those ``states'' necessarily has negative values. 
Indeed, consider the values of $Q$ for $S^{\perp}$. With 90\% probability we have that $X_a=0$ and with the same probability we have that $X_b=0$, but somehow
$X_a+X_b=1$ with 80\% probability. It is easy to verify this can happen only
if we assume a negative probability for $P(1,1)$. Indeed, using the relation (\ref{QtoP}) between $Q$ and $P$ we get $P(1,1)=-0.3$.
This is the point, then, where we have crucially deviated from Spekkens' toy model. We do have valid instrumental states but there are no longer  valid epistemic states corresponding to $S^\perp$ and $S'^\perp$. 

Neither of the special cases $r=0$ or $r=1$ corresponds to the exact violation $B_{QM}=2\sqrt{2}$ allowed by quantum mechanics.
However, we can go one step further. The states $S$ and $S'$ that we defined correspond to valid pure epistemic states, with all probabilities $1\geq P(x_a,x_b)\geq 0$, and only $S^{\perp}$ and $S'^\perp$ do not. But once we have taken that step we may as well
define measurements that make use of {\em two} negative-probability states, $\tilde{S}$ and $\tilde{S}^\perp$.
In that case, though, we have in general a problem calculating the measure $M_r(P)$. Indeed, only for odd integer values of $r$ can we make sense of
$P(x)^{r+1}$ for negative values of $P(x)$. In particular then, let us take $r=1$, and consider any pure states as defined by $N_1(\{Q_i\})=1/2$, including those with negative probabilities $P$. As is easily verified, one obtains the maximum violation of the Bell-CHSH inequality by using the following two states to define measurements:
\begin{eqnarray}
\tilde{S}:Q_0(1)&=&1/2+\sqrt{2}/4\nonumber\\
Q_1(1)&=&1/2+\sqrt{2}/4\nonumber\\
Q_2(0)&=&1/2\nonumber\\
r&=&1,
\end{eqnarray}
and
\begin{eqnarray}
\tilde{S}':
Q_0(1)&=&1/2+\sqrt{2}/4\nonumber\\
Q_1(1)&=&1/2-\sqrt{2}/4\nonumber\\
Q_2(0)&=&1/2\nonumber\\
r&=&1,
\end{eqnarray}
while using the same quantity $B$ of Eq.~(\ref{B}).
These states and their orthogonal complements do correspond to pure states according to the measure $N_1$, but all 4 states have some negative probabilities $P$. The violation of the Bell-CHSH inequality with these measurements is $B_1=2\sqrt{2}$, identical to the quantum-mechanical value $B_{QM}$. This observation, then, provides an additional reason for using $r=1$ as a measure of information useful for quantum mechanics \cite{brukner}.

The question has been posed \cite{pr} why quantum mechanics, without admitting signaling,  allows a finite violation of the Bell-CHSH inequalities but not the maximum. The present toy model has the same property as quantum mechanics. The value $r=1$ is the smallest value compatible with the knowledge balance principle even when allowing negative probabilities, and it leads to the largest violation. The models with other odd integer values for $r$ violate Bell-CHSH inequalities by smaller amounts.
\section{Discussion and conclusions}
We extended the toy model presented by Spekkens in Ref.~\cite{spekkens} by generalizing both his definition of pure states (states of maximal knowledge) and his knowledge balance principle. Our extended model gets even closer to quantum mechanics than the Spekkens model already does.
In particular, although the Spekkens model does not violate Bell-CHSH inequalities, ours does.
In fact, we can get the correct maximum violation of the Bell-CHSH inequalities
by using a particular measure of information; namely the measure of information advocated by Brukner and Zeilinger as being relevant for quantum mechanics,
rather than the Shannon entropy. Perhaps our toy model thus sheds some light on the discussions in Refs.~\cite{brukner,timpson}
about the roles various measures of information could or should play in quantum mechanics. 

Our toy model violates Bell-CHSH inequalities not by giving up locality (in the sense that quantum mechanics itself is a local theory: it is only hidden-variable models mimicking quantum mechanics that are nonlocal) but by allowing negative probabilities for the underlying epistemic states that define the physical states of the toy model. The instrumental states that can be constructed by writing down probabilities of certain fiducial measurements are still characterized by valid (non-negative) probability distributions.
That negative probabilitities, in the form of negative values of an appropriate Wigner function, may be used to indicate or explain nonclassical features has been known for a long time \cite{feynman,scully}.
The probability distributions used in our toy model are not quite the same as discrete Wigner functions for finite-dimensional quantum systems, although they certainly do have many features in common (see, e.g., \cite{wootters, wootters2} and references therein). In particular, those Wigner functions are derived from quantum mechanics, whereas the toy model we considered here is different from quantum mechanics, as was already shown by Spekkens in \cite{spekkens}. In particular, there are 4 toy-model states similar to the 4 Bell states [maximally entangled state of two qubits]. But the toy-model states display different types of correlations and/or anti-correlations than do the Bell states. For example, whereas no quantum-mechanical state of two qubits displays perfect correlations between three mutually unbiased observables, the toy model pure state characterized by $X_a^{(1)}=X_a^{(2)}$
and $X_b^{(1)}=X_b^{(2)}$ does.
It would nevertheless be interesting to study the precise relations between the discrete Wigner functions of \cite{wootters} or of \cite{wootters2} in particular, and the probabilities $P$ and $Q$ defined here.

The fact that the toy model comes close to quantum mechanics but is not quantum mechanics is a good property, we would argue. We should be less surprised if we reproduce many quantum-mechanical features from a ``toy model'' that starts out with, say, complex Hilbert spaces, a tensor product structure, and Hermitian operators. But a smaller surprise gives us less information.

The agreement of the toy model with many features of quantum mechanics once more indicates the importance of the concept of information and states of knowledge
in the quest to understand quantum mechanics \cite{spekkens,zeilinger,fuchs}.
Some open questions still remain about the toy model: Does the Kochen-Specker theorem apply to the toy model? That is, are the measurements we considered contextual? A different type of question is: Can one define the equivalent of a SIC-POVM \cite{sic} in the toy model?
\section*{Appendix}
Here we present a graphical representation of the toy model, similar to the one used by Spekkens in \cite{spekkens}.
An elementary system of type 2 consists of a box with 4 compartments. A state of reality, an ontic state, is a state where one and only one of the compartments is filled. For example, see Figure 2.

A physical state corresponds to an epistemic state, in which our knowledge about which compartment is filled is limited. In a canonical pure state we know that one of two compartments is filled, but we have no idea which one of the two. An example is given in Figure 3.
\begin{figure}
\includegraphics[width=2.5in]{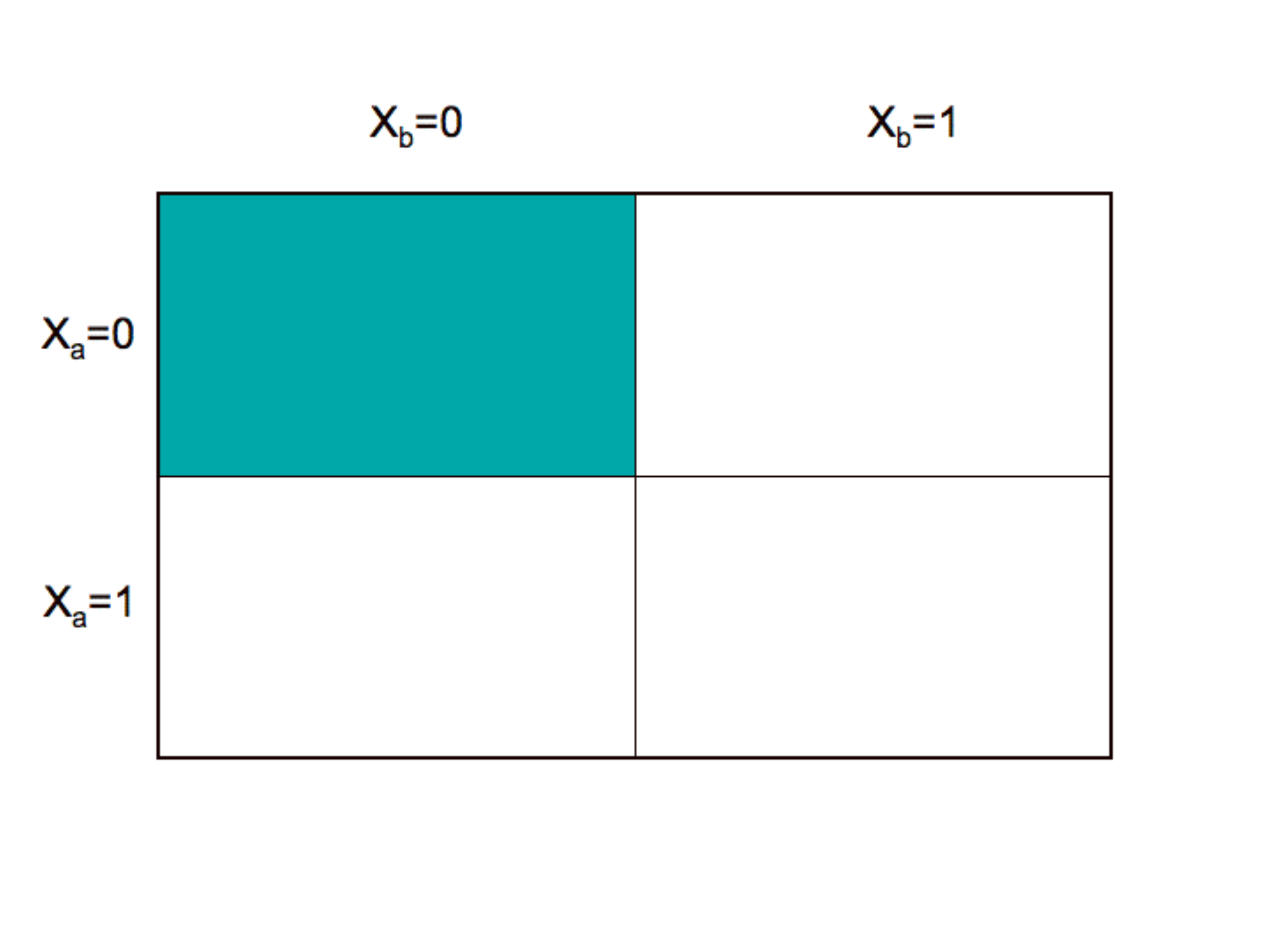}
\caption{An example of an ontic state, in which one particular compartment of the box is filled. Corresponds to a state $X_a=X_b=0$.}
\end{figure}

\begin{figure}
\includegraphics[width=2.5in]{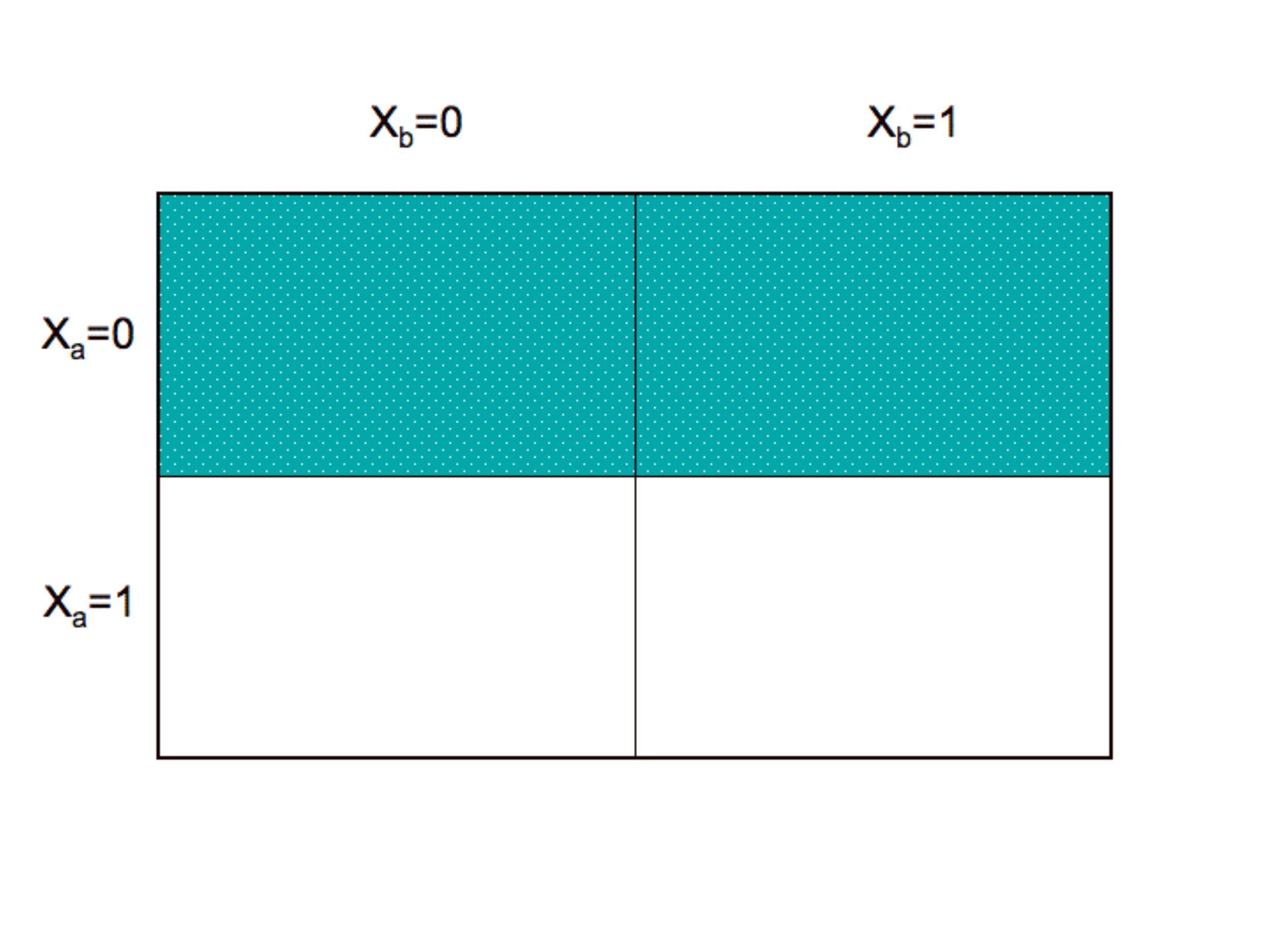}
\caption{An example of an epistemic state, in which we know that exactly one of two particular compartments of the box is filled, but we do not know which one. Corresponds to the state $X_a=0$.}
\end{figure}
We associate with each pure state a measurement or an observable, i.e. a binary question that we can ask the system.
We can ask the system whether one of two particular compartments are filled or not. An example of such a measurement is depicted in Figure 4. The depicted observable corresponds to the algebraic question, ``is $X_a=0$'' or not? That is, it corresponds to the observable $X_0$. 
\begin{figure}
\includegraphics[width=2.5in]{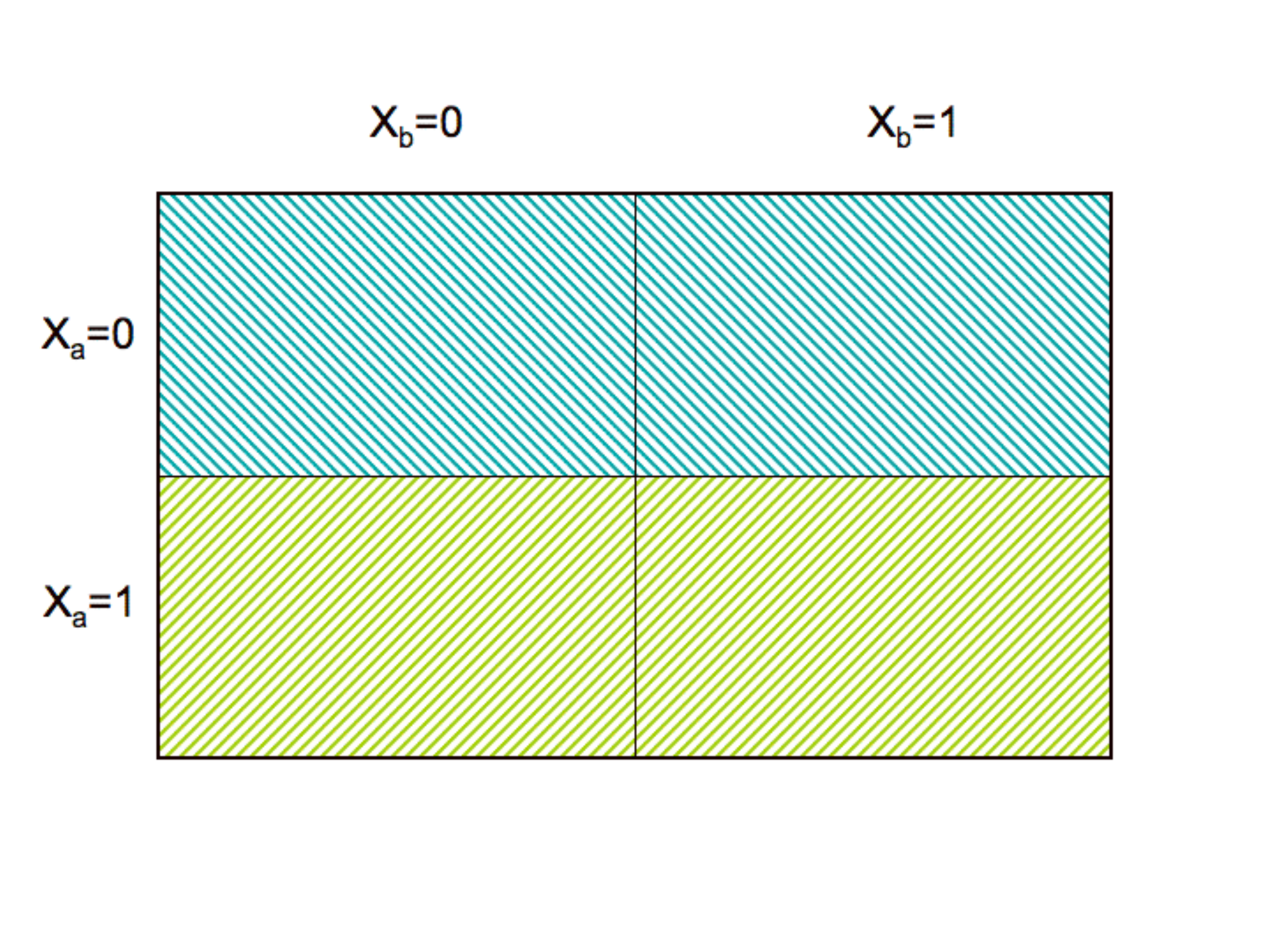}
\caption{An example of an allowed measurement, associated with the pure state of Figure 3. We can ask whether one of the upper two compartments is filled or not. This corresponds to the question whether $X_a=0$ or not.}
\end{figure}
The most general epistemic state allowed is one where we assign 4 probabilities $P(x_a,x_b)$, such that a suitable measure of information $M_r$ (defined in Eq.~(\ref{M})) has a particular value, namely, $M_r(P)=1/2$ for pure states
and $M_r(P)<1/2$ for all other states. When we allow the probabilities $P$ to become negative for some entries, but without allowing negative probabilities for measurement outcomes, we can violate Bell-CHSH inequalities.
\begin{figure}
\includegraphics[width=2.5in]{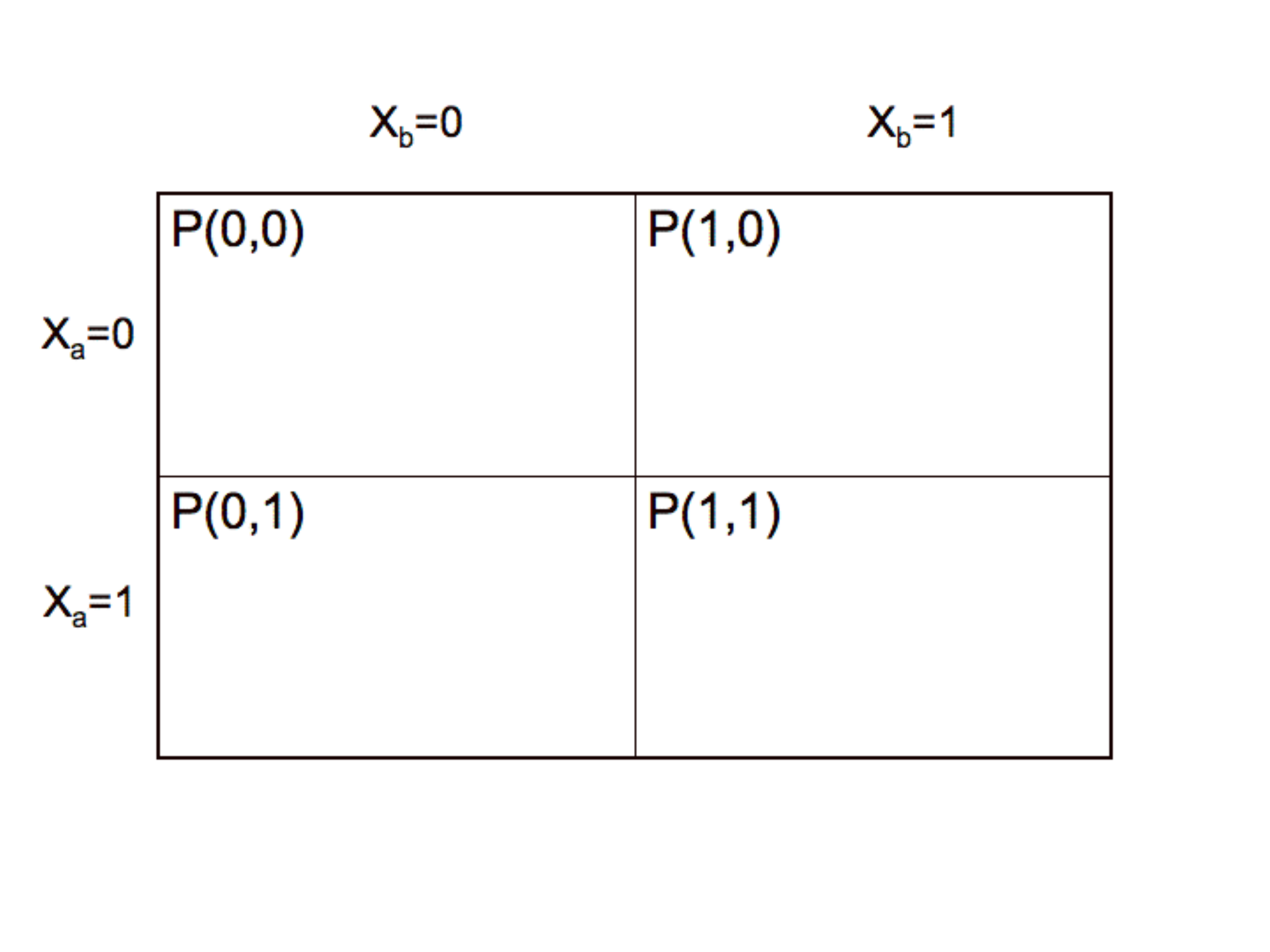}
\caption{An example of a more general epistemic state, in which we have assigned some probability for each compartment to be filled.
$P$ satisfies $\sum_{ij}P(i,j)=1$ and $M_r(P)\leq 1/2$.
Moreover, the two entries in one column, or in one row, or in one diagonal, must sum to a positive number. In the extended toy model, {\em one} entry may be negative.
}
\end{figure}
 
\end{document}